\documentclass[aps,prc,reprint,superscriptaddress,floatfix,amsmath,hyperref,nofootinbib]{revtex4-2}
\usepackage{graphicx,epstopdf}
\usepackage{graphicx,color, subfigure}
\usepackage{booktabs}
\usepackage{multirow}

\usepackage{xcolor}

\begin{document}

\title{Mass measurements of neutron-rich nuclides using the Canadian Penning Trap to inform predictions in the $r$-process rare-earth peak region}

\author{D. Ray}
\email[Corresponding author: ]{dray@triumf.ca}
\altaffiliation[Present address: ]{TRIUMF, 4004 Wesbrook Mall, Vancouver, BC V6T 2A3, Canada}
\affiliation{Department of Physics and Astronomy, University of Manitoba, Winnipeg,  MB, R3T 2N2, Canada}
\affiliation{Physics Division, Argonne National Laboratory, Lemont, IL 60439, USA}

\author{N. Vassh}
\email[Corresponding author: ]{nvassh@triumf.ca}
\affiliation{TRIUMF, 4004 Wesbrook Mall, Vancouver, BC V6T 2A3, Canada}

\author{B. Liu}
\affiliation{Department of Physics and Astronomy, University of Notre Dame, Notre Dame, IN 46556, USA}
\affiliation{Physics Division, Argonne National Laboratory, Lemont, IL 60439, USA}

\author{A.A. Valverde}
\affiliation{Department of Physics and Astronomy, University of Manitoba, Winnipeg,  MB, R3T 2N2, Canada}
\affiliation{Physics Division, Argonne National Laboratory, Lemont, IL 60439, USA}

\author{M. Brodeur}
\affiliation{Department of Physics and Astronomy, University of Notre Dame, Notre Dame, IN 46556, USA}

\author{J.A. Clark}
\affiliation{Physics Division, Argonne National Laboratory, Lemont, IL 60439, USA}
\affiliation{Department of Physics and Astronomy, University of Manitoba, Winnipeg,  MB, R3T 2N2, Canada}

\author{G.C. McLaughlin}
\affiliation{Department of Physics, North Carolina State University, Raleigh, North Carolina 27695, USA}

\author{M.R. Mumpower}
\affiliation{Theoretical Division, Los Alamos National Laboratory, Los Alamos, New Mexico 87545, USA}

\author{R. Orford}
\affiliation{Nuclear Science Division, Lawrence Berkeley National Laboratory, Berkeley, CA 94720, USA}

\author{W.S. Porter}
\affiliation{Department of Physics and Astronomy, University of Notre Dame, Notre Dame, IN 46556, USA}

\author{G. Savard}
\affiliation{Physics Division, Argonne National Laboratory, Lemont, IL 60439, USA}
\affiliation{Department of Physics, University of Chicago, Chicago, IL 60637, USA}

\author{K. S. Sharma}
\affiliation{Department of Physics and Astronomy, University of Manitoba, Winnipeg,  MB, R3T 2N2, Canada}

\author{R. Surman}
\affiliation{Department of Physics and Astronomy, University of Notre Dame, Notre Dame, IN 46556, USA}

\author{F. Buchinger}
\affiliation{Department of Physics, McGill University, Montreal, QC H3A 2T8, Canada}

\author{D.P. Burdette}
\affiliation{Physics Division, Argonne National Laboratory, Lemont, IL 60439, USA}

\author{N. Callahan}
\affiliation{Physics Division, Argonne National Laboratory, Lemont, IL 60439, USA}

\author{A.T. Gallant}
\affiliation{Nuclear and Chemical Sciences Division, Lawrence Livermore National Laboratory, Livermore, CA 94550, USA}

\author{D.E.M. Hoff}
\affiliation{Nuclear and Chemical Sciences Division, Lawrence Livermore National Laboratory, Livermore, CA 94550, USA}

\author{K. Kolos}
\affiliation{Nuclear and Chemical Sciences Division, Lawrence Livermore National Laboratory, Livermore, CA 94550, USA}

\author{F.G. Kondev}
\affiliation{Physics Division, Argonne National Laboratory, Lemont, IL 60439, USA}

\author{G. E. Morgan}
\affiliation{Physics Division, Argonne National Laboratory, Lemont, IL 60439, USA}
\affiliation{Department of Physics and Astronomy, Louisiana State University, Baton Rouge, LA 70803, USA}

\author{F. Rivero}
\affiliation{Department of Physics and Astronomy, University of Notre Dame, Notre Dame, IN 46556, USA}

\author{D. Santiago-Gonzalez}
\affiliation{Physics Division, Argonne National Laboratory, Lemont, IL 60439, USA}

\author{N.D. Scielzo}
\affiliation{Nuclear and Chemical Sciences Division, Lawrence Livermore National Laboratory, Livermore, CA 94550, USA}

\author{L. Varriano}
\altaffiliation[Present address: ]{Center for Experimental Nuclear Physics and Astrophysics, University of Washington, Seattle, WA 98195, USA}
\affiliation{Physics Division, Argonne National Laboratory, Lemont, IL 60439, USA}
\affiliation{Department of Physics, University of Chicago, Chicago, IL 60637, USA}

\author{C.M. Weber}
\affiliation{School of Physics and Astronomy, University of Minnesota Twin Cities, MN, Minnesota 55455, USA}

\author{G. E. Wilson}
\affiliation{Department of Physics and Astronomy, Louisiana State University, Baton Rouge, LA 70803, USA}
\affiliation{Physics Division, Argonne National Laboratory, Lemont, IL 60439, USA}

\author{X.L. Yan}
\affiliation{Institute of Modern Physics, Chinese Academy of Sciences, Lanzhou 730000, China}

\begin{abstract}

Studies aiming to determine the astrophysical origins of nuclei produced by the rapid neutron capture process ($r$ process) rely on nuclear properties as inputs for simulations. The solar abundances can be used as a benchmark for such calculations, with the $r$-process rare-earth peak (REP) around mass number ($A$) 164 being of special interest due to its presently unknown origin. With the advancement of rare isotope beam production over the last decade and improvement in experimental sensitivities, many of these REP nuclides have become accessible for measurement. Masses are one of the most critical inputs as they impact multiple nuclear properties, namely the neutron-separation energies, neutron capture rates, $\beta$-decay rates, and $\beta$-delayed neutron emission probabilities. In this work, we report masses of 20 neutron-rich nuclides (along the Ba, La, Ce, Pr, Nd, Pm, Gd, Dy and Ho isotopic chains) produced at the CAlifornium Rare Isotope Breeder Upgrade (CARIBU) facility at Argonne National Laboratory. The masses were measured with the Canadian Penning trap (CPT) mass spectrometer using the Phase-Imaging Ion-Cyclotron-Resonance (PI-ICR) technique. We then use these new masses along with previously published CPT masses to inform predictions for a Markov Chain Monte Carlo (MCMC) procedure aiming to identify the astrophysical conditions consistent with both solar data and mass measurements. We show that the MCMC responds to this updated mass information, producing refined results for both mass predictions and REP abundances.
\end{abstract}

\maketitle


\section{Introduction}\label{sec:intro}

Looking at the relative abundance of elements in the solar system for hints of the astrophysical conditions that led to the formation of elements heavier than iron has been central to nuclear astrophysics from its very beginning~\cite{B2FH}. 
Synthesis of about half of these elements is attributed to the rapid neutron capture process ($r$ process), of which our knowledge is still limited.
Early studies indicated the significance of structures or `peaks' in the abundances, corresponding to stable neutron-rich species like those at neutron shell closures.  
Over the last decade, this line of investigation has led to experimental campaigns around the world pursuing measurements of neutron-rich systems to probe the nuclear properties responsible for these structures seen in abundances.

A feature that is especially interesting to study is that of the rare-earth peak around mass number $A=164$ seen in the solar abundances, which is tied to the $r$ process.
Since the $r$ process involves nuclei far more neutron-rich than other astrophysical processes, nuclear data of highly unstable short-lived systems are needed to inform calculations. 
Atomic masses are an extremely important piece of data as shown by sensitivity studies~\cite{mumpower_impact_indi_nuc_prop, mumpower_impact_indi_masses, brett_sensitivity}.
Particularly over the last decade, mass measurements of neutron-rich isotopes of rare-earth elements La (proton number $Z=57$), Ce ($Z=58$), Pr ($Z=59$), Nd ($Z=60$), Pm ($Z=61$), Sm ($Z=62$), Eu ($Z=63$), and Gd ($Z=64$) have entered the regime of interest for rare-earth peak formation~\cite{orford_prl_Nd-Sm_2018,orford_prc_isomers_2020,thesis_cpt_orford_phd_long, orford_2021_rare_earth2_publshd}.
The masses of these nuclides, measured with the Canadian Penning Trap mass spectrometer (CPT)~\cite{guy_cpt_first_1997,guy_cpt_anl_first_2001} have been used
to constrain the $r$-process conditions that can reproduce the peak~\cite{VasshMCMC}. 
However, a powerful aspect of these measurements that is not captured by individual 
impact studies is their influence when considered as a full suite of new information. 
In this work,  the mass values are incorporated into a Markov chain Monte Carlo (MCMC) procedure (outlined in Ref.~\cite{VasshMCMC}) to see how its mass predictions respond when informed with the complete set of $Z=57-64$ measurements conducted by the CPT over the last decade.

The experimental methods and latest mass measurement results of neutron-rich isotopes of elements Ba, La, Ce, Pr, Nd, Pm, Gd, Dy and Ho, including first measurements of $^{152}$La and $^{168}$Gd,
are discussed first.  Next, the mass values are applied in an $r$-process abundance impact study to observe their direct influence on predictions for the $r$-process rare-earth peak. 
Lastly,  the sensitivity of the MCMC method to the suite of measurements achieved by the CPT campaign is demonstrated by showing how this data refine the MCMC mass predictions in the regions beyond current measurements.
Thus, both the direct and indirect impact on abundances that can be achieved with measurements along neighboring isotopic chains are highlighted.

\section{Experimental Method}\label{sec:expt_method}

The nuclides were produced at the CAlifornium Rare Isotope Breeder Upgrade (CARIBU) facility~\cite{guy_caribu_short_1_2008} at Argonne National Laboratory using a $\sim0.5$~Ci $^{252}$Cf source, which spontaneously decays with a fission branch of 3.1028(7)\%~\cite{nubase2020_Kondev2021}.
The fission fragments were stopped, cooled and extracted in a large volume gas catcher~\cite{guy_caribu_gas_catcher_short_1_2016} using high purity helium gas and a combination of radiofrequency (RF) and direct current (DC) fields.
The extracted ions, belonging to multiple species with $A$ ranging from $80$ to $180$ and charge states ($q$) of primarily $1+$ or $2+$, underwent a first stage of selection by passing through an isobar separator~\cite{caribu_isobar_separator_1_2008}, consisting of a pair of dipole magnets tuned to select ions of a selected $A/q$. 
The continuous stream of ions then passed through an RF quadrupole cooler-buncher, where their energy spread was reduced and they were converted into a bunched beam released every $50$ ms.
The bunched beam was further mass resolved using a multi-reflection time-of-flight (MR-TOF) mass separator~\cite{hirsh_caribu_mrtof}.
For the current measurements, the ions were held inside the \mbox{MR-TOF} for $15-20$ ms enabling achievement of a mass resolution that can reach $R = m / \Delta m \approx 100$,$000$, enough to separate ions of different isobaric species in the beam.
At the exit of the \mbox{MR-TOF}, the mass-resolved bunched beam passed through a Bradbury-Nielsen gate where the nuclides of interest were selected. These selected ions were then sent to the CARIBU low-energy experimental area, where they would undergo further cooling inside a linear Paul trap before being injected into the CPT.

\begin{figure}[ht]
\centering
 \includegraphics[width=0.95 \columnwidth]{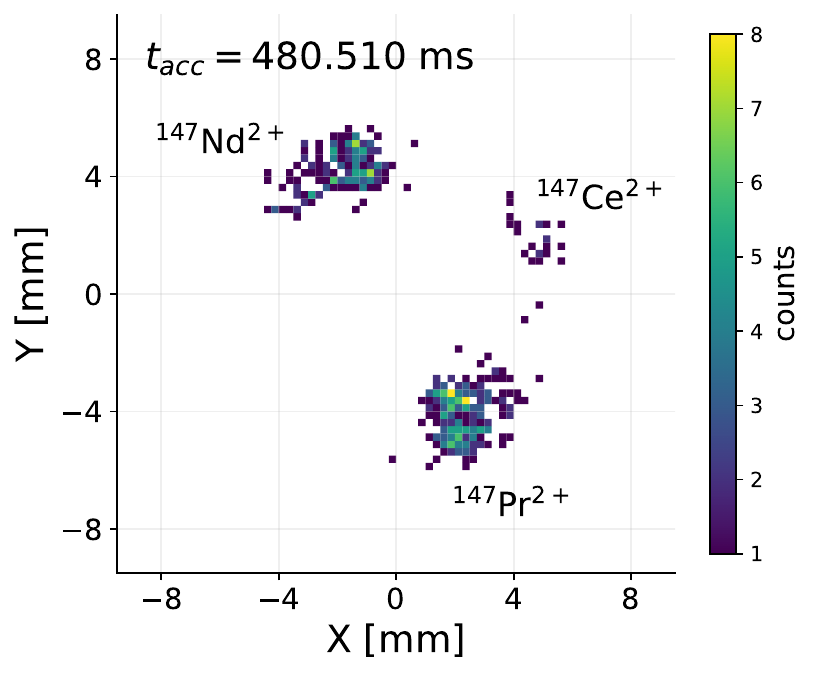}
 \caption{Position spectrum of a final phase measurement of $^{147}$Nd$^{2+}$ ions with $t_{\text{acc}}=480.510$ ms. Also shown are spots of two isobaric species present in the beam: $^{147}$Pr$^{2+}$ and $^{147}$Ce$^{2+}$.\label{fig:147Nd_xy}}
\end{figure}

Once captured and trapped in the CPT, the cyclotron frequency ($\nu_{c}$) of the ions was measured using the Phase-Imaging Ion-Cyclotron-Resonance (PI-ICR) technique~\cite{eliseev_piicr_prl_2013}.
A detailed description of the measurement procedure at the CPT can be found in Ref.~\cite{orford_nimb_piicr_cpt_2020,thesis_cpt_ray_phd_long}. 
In short, it consists of simultaneous measurement of the ions' magnetron ($\nu_{-}$) and reduced cyclotron ($\nu_{+}$) radial eigenmotions. 
For each measurement, the ions first accumulate a $\nu_{+}$ phase over some accumulation time $t_{\text{acc}}$. At that point a $\nu_{c}$ excitation is applied converting their $\nu_{+}$ motion to $\nu_{-}$ motion.
They next accumulate a $\nu_{-}$ phase over time $T-t_{\text{acc}}$ before exiting the trap, where $T$ is the total measurement time in the trap excluding the durations of different RF excitations.
Two measurements are conducted, a reference phase measurement where $t_{\text{acc}}=0$ with the observed phase mostly from $\nu_{-}$ motion, and a final phase measurement where $t_{\text{acc}}>0$ with both motions contributing to the observed phase. 
By measuring the phase difference $\phi_{c}$, $\nu_{c}$ can be determined as:
\begin{equation}\label{eq:wc}
\nu_{c} = \frac{\phi_{c} + 2 \pi \mathcal{N}}{2 \pi t_{\text{acc}}} ,
\end{equation}
where $\mathcal{N}$ is the total number of complete cyclotron revolutions a target-nuclide ion completes in time $t_{\text{acc}}$.
Initial measurements varying $t_{\text{acc}}$ from single to hundreds of ms are conducted to identify all the species present in the beam. A longer $t_{\text{acc}}$ is then selected for precision measurement such that all the beam species get resolved and manifest as separate final spots. This can be seen in Fig.~\ref{fig:147Nd_xy}, where the measurement of $^{147}$Nd was being conducted at $t_{\text{acc}}=480.510$ ms using a beam that also contained ions of other isobaric species namely $^{147}$Pr and $^{147}$Ce.
A simple mean-shift algorithm \cite{meanshift} was used to cluster the data before performing a Gaussian fit to find the spot-centroids and conducting further analysis.

\begin{figure}[ht]
\centering
 \includegraphics[width=0.9 \columnwidth]{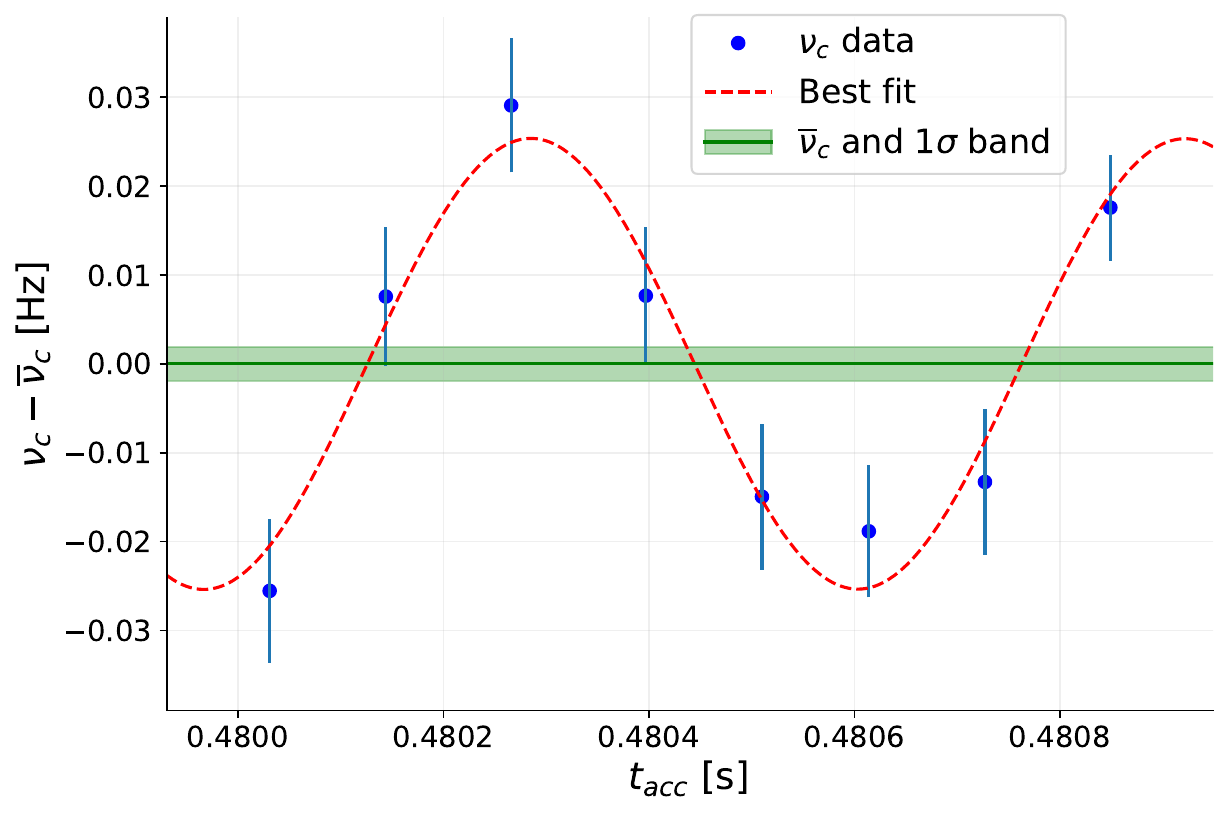}
 \caption{Measured $\nu_{c}$ values for $^{147}$Nd$^{2+}$ ions using eight distinct $t_{\text{acc}}$ values between
480.0 ms and 480.9 ms. The red dashed line represents a fit~\cite{orford_nimb_piicr_cpt_2020} of the data. The green horizontal line and bar shows the true $\overline{\nu}_{c}$. \label{fig:147Nd_sine}}
\end{figure}

The biggest systematic effect observed at the CPT arises from the residual $\nu_{-}$ motion the ions possess at the onset of the $\nu_{+}$ excitation~\cite{orford_nimb_piicr_cpt_2020}. This motion carries over to the detected final phase, resulting in a sinusoidal dependence of the measured $\nu_{c}$ on the $t_{\text{acc}}$ around the true cyclotron frequency ($\overline{\nu}_{c}$). 
To accurately correct for this effect, a series of $\nu_{c}$ measurements are conducted by varying $t_{\text{acc}}$ over at least one $\nu_{-}$ period, and are then fit to the model described in Ref.~\cite{orford_nimb_piicr_cpt_2020} to obtain $\overline{\nu}_{c}$.
Figure~\ref{fig:147Nd_sine} shows this sinusoidal behavior observed during the measurement of $^{147}$Nd.
Another systematic effect that is considered during analysis is a shift in the true reference phase due to the non-zero phase advance of the isobaric contaminants in the trap during the short excitation pulses. This was corrected following the procedure in Ref.~\cite{orford_nimb_piicr_cpt_2020}.
The strength of the magnetic field is determined by measuring $\nu_{c}$ of a calibrant (Cal) species, with $\nu_{c}$ close to the target nuclide and having a well-known mass, around the same $t_{\text{acc}}$.
Next, the ratio of these two $\overline{\nu}_{c}$ is determined: $r = \overline{\nu}_{c,\text{Cal}} / \overline{\nu}_{c}$. 
A systematic shift from differences in the measurement angles between the target and calibrant nuclides was accounted for by applying a correction to $r$ as described in Ref.~\cite{liu2024precisemassmeasurement108}. To take a cautious approach, the same shift was also added in quadrature to the uncertainties.
Effects related to the difference in $A/q$ between the calibrant and measured nuclide have been determined to be less than 4.1~ppb$/$u from measurements of isotopes of well-known masses~\cite{liu2024precisemassmeasurement108}. This shift was applied to the measured masses, and also added in quadrature to the final uncertainties to be conservative. 
Additional systematic effects including 
magnetic field drift, 
ion-ion interactions, 
non-circular projection from the trap to the detector and electric field instabilities have been studied, and observed to have a cumulative effect of less-than $3.2$~ppb~\cite{thesis_cpt_ray_phd_long, ray2024phaseimagingioncyclotronresonancemassspectrometry}. This was added in quadrature to the uncertainties of the measurements.
The mass ($M$) is determined from:
\begin{equation}\label{eq:mass}
M  =  \displaystyle r \frac{q}{q_{\text{Cal}}} (M_{\text{Cal}} - q_{\text{Cal}} m_{e}) + q m_{e},
\end{equation}
where $m_{e}$ is the mass of an electron. The electron binding energies for the isotopes presented here are over an order of magnitude smaller than the achieved precision in measured masses, and were therefore ignored.

\section{Results}\label{sec:label}

\begin{figure*}
\centering
\includegraphics[width = 0.85\textwidth]
{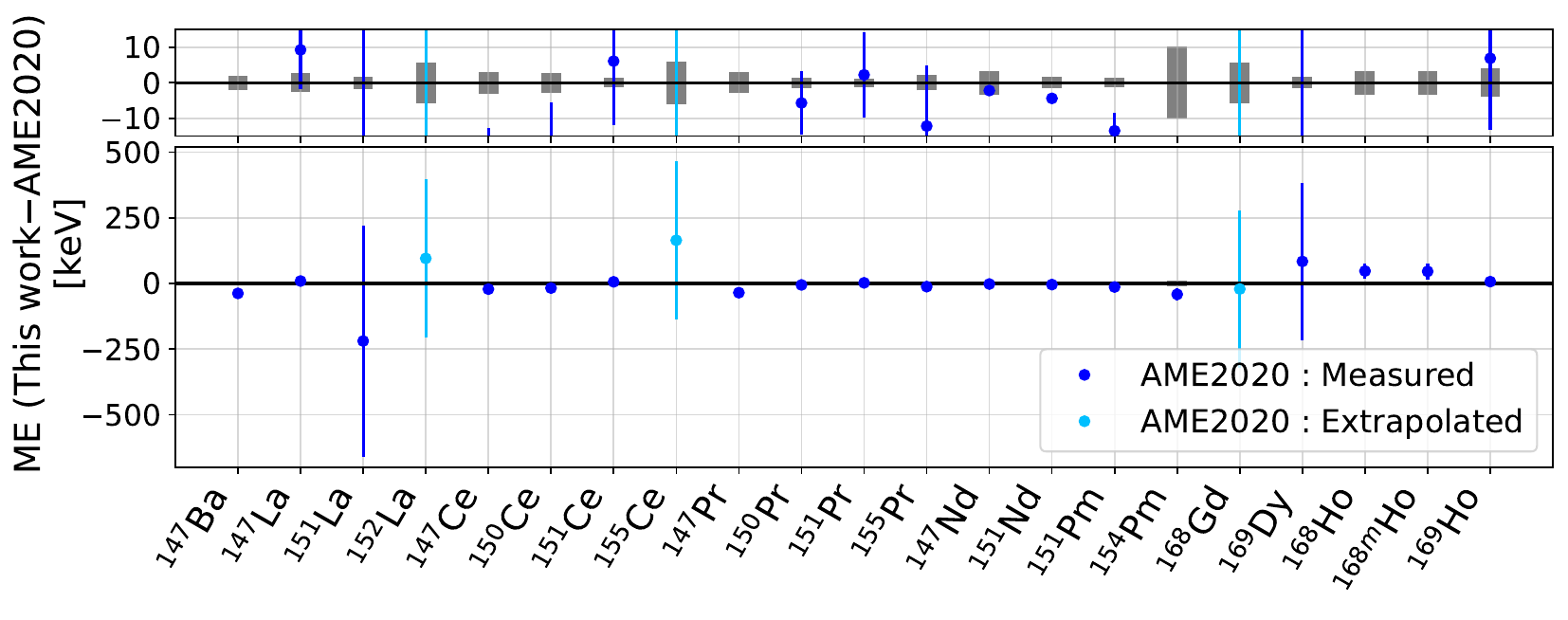}
\caption{Comparison between mass excess (ME) from this work and the AME2020~\cite{ame2020_b_Wang2021, nubase2020_Kondev2021}. The upper panel shows the comparisons, zoomed in on the results from this work to highlight the precisions presented here, shown by the grey bars.
\label{fig:comparison_results_ame2020}}
\end{figure*}

Masses of 20 isotopes 
are presented in this work, including first reported masses of $^{152}$La and $^{168}$Gd, and first direct mass measurement of an isomer of $^{168}$Ho.
Additionally, 
$^{155}$Ce and $^{168}$Gd, produced from a $10^{-5}$\% branch of $^{252}$Cf fission, and $^{152}$La, $^{169}$Dy and $^{168,169}$Ho, populated from a $10^{-4}$ branch,  comprise some of the most weakly produced isotopes from $^{252}$Cf-fission whose masses have been measured. A summary of the results, along with those published in the latest version of the Atomic Mass Evaluation (AME2020)~\cite{ame2020_b_Wang2021, nubase2020_Kondev2021} is shown in Table~\ref{tab:mass}, while a comparison between the two is shown in Fig.~\ref{fig:comparison_results_ame2020}. Also included in Table~\ref{tab:mass} are results from other literature that were reported after the publication of the AME2020.

\begin{table*}[ht!]
 \caption{\label{tab:mass} Summary of measurements presented in this work, including the isotopes whose masses were measured, their half-lives, the calibrant ions used, the $\overline{\nu}_{c}$ ratio $r$, and mass excess values from this work and from literature.}
\begin{ruledtabular}
\begin{tabular}{c c c c c c c}
       Target & Half-life& Calibrant & $\overline{\nu}_{c}$ ratio & \multicolumn{3}{c}{Atomic Mass Excess [keV]} \\
    \cmidrule(lr){5-7} 
    Ion & $t_{1/2}$~\cite{nubase2020_Kondev2021} & Ion & $r = \overline{\nu}_{c, \text{Cal}} / \overline{\nu}_{c}$ & CPT (this work) & AME2020~\cite{ame2020_b_Wang2021, nubase2020_Kondev2021} & Other literature \\ \hline

$^{147}$Ba$^{2+}$  & 893(1)~ms & C$_{6}$H$_{6}^{+}$ & 0.941 325 693(14) & -60301.9(20) & -60264(20) & - \\ [5pt]

$^{147}$La$^{2+}$  & 4.026(2)~s & C$_{6}$H$_{6}^{+}$ & 0.941 281 904(18) & -66668.8(27) & -66678(11) & - \\ 
$^{151}$La$^{2+}$  & 465(24)~ms & C$_{6}$H$_{6}^{+}$  & 0.966 998 054(12) & -53529.1(17) & -53310(440) & -53542(17) $^{\text{a}}$ \\ 
$^{152}$La$^{2+}$  & 287(16)~ms & C$_{6}$H$_{6}^{+}$  & 0.973 434 310(40) & -49194.7(58) & -49290(300)\# & - \\ [5pt]

$^{147}$Ce$^{2+}$  & 56.4(1)~s & C$_{6}$H$_{6}^{+}$  &0.941 244 992(21)  & -72035.7(30)  & -72014(9) & - \\ 
$^{150}$Ce$^{2+}$  & 6.05(7)~s & C$_{6}$H$_{6}^{+}$  & 0.960 513 648(20) & -64864.5(29) & -64847(12) & - \\ 
$^{151}$Ce$^{2+}$  & 1.76(6)~s & C$_{6}$H$_{6}^{+}$  & 0.966 945 166 6(89) & -61218.9(13) & -61225(18)& -61230(15) $^{\text{a}}$ \\ 
$^{155}$Ce$^{2+}$  & 313(7)~ms& C$_{6}$H$_{6}^{+}$  & 0.992 664 505(41) & -47616(6) & -47780(300)\# & -47576(31) $^{\text{a}}$ \\  [5pt]

$^{147}$Pr$^{2+}$  & 13.39(4)~m & C$_{6}$H$_{6}^{+}$  & 0.941 221 309(20) & -75479.2(29) & -75444(16) & - \\ 
$^{150}$Pr$^{2+}$  & 6.19 (16)~s & $^{150}$Nd$^{2+}$   & 1.000 038 476 9(59) & -68306.7(14) & -68301(9) & - \\ 
$^{151}$Pr$^{2+}$  & 18.90 (7)~s & C$_{6}$H$_{6}^{+}$   & 0.966 906 935 0(83) & -66777.8(12) & -66780(12) & - \\ 
$^{155}$Pr$^{2+}$  & 1.47 (3)~s & C$_{6}$H$_{6}^{+}$  & 0.992 610 781(15) & -55427.2(21) & -55415(17) & -55431(14) $^{\text{a}}$ \\  [5pt]

$^{147}$Nd$^{2+}$  & 10.98 (1)~d & C$_{6}$H$_{6}^{+}$  & 0.941 202 947(23) & -78149.0(33) & -78146.8(13) & - \\ 
$^{151}$Nd$^{2+}$  & 12.44(7)~m & C$_{6}$H$_{6}^{+}$  & 0.966 878 257(11) & -70947.6(16) & -70943.2(11) & -70945(20) $^{\text{a}}$ \\ [5pt]

$^{151}$Pm$^{2+}$  & 28.40 (4)~h & C$_{6}$H$_{6}^{+}$  & 0.966 861 393 6(95) & -73399.5(14) & -73386(5) & -73386(24) $^{\text{a}}$ \\ 
$^{154}$Pm$^{2+}$ $^{\text{b}}$ & 2.68(7)~m & $^{154}$Nd$^{2+}$  & 0.999 980 968(71)  & -68308(10) & -68267(25) & -68303(27) $^{\text{a}}$ \\  [5pt]

$^{168}$Gd$^{2+}$  & 3.03(16)~s & $^{84}$Kr$^{+}$ & 1.000 746 569(36) & -48171.2(56) & -48150(300)\# & - \\  [5pt]

$^{169}$Dy$^{2+}$  & 39(8)~s & $^{84}$Kr$^{+}$  & 1.006 658 281(10) & -55516.3(16) & -55600(300) & -55523.1(57) $^{\text{c}}$ \\  [5pt]

$^{168}$Ho$^{2+}$  & 2.99(7)~m & $^{84}$Kr$^{+}$  & 1.000 670 819(21) & -60012.8(33) & -60060(30) & - \\ 
$^{168m}$Ho$^{2+}$  & 132(4)~s & $^{84}$Kr$^{+}$  & 1.000 671 194(21) & -59954.2(33) & -60000(30) & - \\ 
$^{169}$Ho$^{2+}$  & 4.72 (1)~m & $^{85}$Rb$^{+}$  & 0.994 778 711(25) & -58789.1(40) & -58796(20) & -58820.7(47) $^{\text{c}}$ \\ [5pt]

\end{tabular}
\end{ruledtabular}
$^{\text{(a)}}$ Ref. \cite{kimura_PhysRevC.110.045810_2024} \hspace{1mm}
$^{\text{(b)}}$ Uncertainty inflated due to possible mixture of ground and isomeric states. See text for details. \hspace{1mm}
$^{\text{(c)}}$ Ref. \cite{jaries2024probing}
\end{table*}

\subsection{Individual nuclides}\label{subsec:indi_nucl}

\noindent \underline{$A = 147: \; ^{147}$Ba, $^{147}$La, $^{147}$Ce, $^{147}$Pr and $^{147}$Nd}
\vspace{2mm}

The mass excess values reported in AME2020~\cite{ame2020_b_Wang2021} for $^{147}$Ba and $^{147}$Ce are mainly based on CPT \mbox{TOF-ICR} measurements from Ref.~\cite{savard_2006_int_j_mass_spec_1}. The current CPT \mbox{PI-ICR} measurements are around 2$\sigma$ heavier with uncertainties smaller by factors of about 10 and 3, respectively.
The $^{147}$La mass excess reported here agrees with the AME2020~\cite{ame2020_b_Wang2021} value, also adapted from Ref.~\cite{savard_2006_int_j_mass_spec_1}, with 4-times smaller errors.
The AME2020~\cite{ame2020_b_Wang2021} mass excess for $^{147}$Pr mass excess is an average of multiple $\beta$-endpoint measurements ~\cite{147Pr_HOFFMAN19641769, 147Pr_YAMAMOTO1981855, 151Nd_101143_JPSJ643244}. 
The present CPT-measured value is 2$\sigma$ heavier, and around 5 times more precise.
The AME2020~\cite{ame2020_b_Wang2021} mass value of $^{147}$Nd is a combination of two $(n,\gamma)$~\cite{147Nd_n_gamma_1_ROUSSILLE1975380, 147Nd_n_gamma_2_firestone2004database} and a $(d,p)$~\cite{147Nd_dp_WIEDNER1967433} measurements.
The CPT-measured mass excess agrees with the AME2020 value.

\vspace{3mm} \noindent \underline{$A = 150: \; ^{150}$Ce, $^{150}$Pr}
\vspace{2mm}

The $^{150}$Ce and $^{150}$Pr masses in the AME2020~\cite{ame2020_b_Wang2021} are also evaluated based on results from Ref.~\cite{savard_2006_int_j_mass_spec_1}. The current results agree with the AME2020, with increased precision of factors around 4 and 7, respectively.

\vspace{3mm} \noindent \underline{$A = 151: \; ^{151}$La, $^{151}$Ce, $^{151}$Pr, $^{151}$Nd, and  $^{151}$Pm}
\vspace{2mm}

The $^{151}$La mass in the AME2020~\cite{ame2020_b_Wang2021} is from measurements done at the FRS-ESR, GSI 
\cite{151La_matos2004isochronous, 151La_knobel2016new2}. The CPT result here agrees with the AME2020, while being 250 times more precise.
The $^{151}$Ce and $^{151}$Pr masses in the AME2020~\cite{ame2020_b_Wang2021} were adapted mainly from Ref.~\cite{savard_2006_int_j_mass_spec_1}.
The results here agree with both, with uncertainties smaller by factors of 14 and 10.
The AME2020~\cite{ame2020_b_Wang2021} $^{151}$Nd mass is an average of $(n,\gamma)$ measurements~\cite{151Nd_PINSTON197661, 147Nd_n_gamma_2_firestone2004database}, while that of $^{151}$Pm is a combination of a $(^{3}\text{He}, d)$~\cite{151Pm_80st10} and a $\beta$-endpoint~\cite{151Pm_BERTELSEN1964657} measurement. The CPT-measured masses for both nuclides are around 2.5$\sigma$ lighter, while the precision for $^{151}$Pm has been improved by a factor of about 4. Since the publication of the AME2020, the masses of $^{151}$La, $^{151}$Nd and $^{151}$Pm have been measured using the \mbox{MR-TOF} mass spectrograph at RIKEN
\cite{kimura_PhysRevC.110.045810_2024}. The CPT results presented here agree with all of them, while reporting uncertainties smaller by factors of about 10, 13 and 17, respectively.
The authors of~\cite{kimura_PhysRevC.110.045810_2024} also reported the mass of $^{151}$Ce, claiming their observed peak could be a mixture of the ground and the isomeric state. 
No isomer is reported in the AME2020~\cite{nubase2020_Kondev2021} or was observed during the CPT measurement. The experiment was conducted with $t_{\text{acc}} \approx 480$~ms, which would be enough to resolve any isomers with excitation energies of $>30$~keV. This result agrees with the reported masses in Ref.~\cite{kimura_PhysRevC.110.045810_2024} with a $\sim$12-times improved precision.

\vspace{3mm} \noindent \underline{$A = 152: \; ^{152}$La}
\vspace{2mm}

This is the first direct measurement of the mass of $^{152}$La, with the AME2020~\cite{ame2020_b_Wang2021} reporting extrapolated values. The measurement precision of $\sim4\times10^{-8}$ achieved at the CPT was limited by $^{152}$La being produced from a $10^{-4}$\% branch of $^{252}$Cf fission~\cite{EnR}, as well as by its short half-life ($t_{1/2}$) of 287(16)~ms~\cite{nubase2020_Kondev2021}.

\vspace{3mm} \noindent \underline{$A = 154: \; ^{154}$Pm}
\vspace{2mm}

The existence of two long-lived states in $^{154}$Pm is well established following early $\beta$-decay spectroscopy studies~\cite{154Pm_d1971isomerism, 154Pm_tannila1972evidence, 154Pm_preiss1973isomerism, 154Pm_yamamoto1974decay}. The NUBASE2020 evaluation~\cite{nubase2020_Kondev2021} associates the longer-lived activity ($t_{1/2}=2.68(7)$~m) with the ground state and the $J^{\pi}=(4+)$ assignment is proposed, while the isomer is assigned $J^{\pi}=(1-)$ and related to the shorter-lived activity ($t_{1/2}=1.73(10)$~m). 
However, the excitation energy (and mass) of the isomer is not well known. The value of $-230(50)$~keV for the excitation energy of the isomer in Ref.~\cite{nubase2020_Kondev2021} is spurious and is based on $\beta$-decay end point measurements~\cite{ame2020_a_Huang2021} that suffer from incomplete knowledge of the decay schemes for the ground state and the isomer. 

Both states belong to a $\sim$0.3\% branch from $^{252}$Cf fission according to literature~\cite{EnR}.
While measuring the mass of $^{154}$Pm, a thorough search for the isomeric state was conducted.
A potential isomer with excitation energy between 10 and 20~keV was observed at certain accumulation times, but due to some technical issues it could not be conclusively identified and measured.
As a result, the uncertainty on the mass of the ground state was inflated considering a possible mixture of equal proportion between the ground state and an isomeric state of excitation energy of 20~keV.
The result agrees within 1.6$\sigma$ of the AME2020 value~\cite{ame2020_b_Wang2021} with an uncertainty smaller by a factor of about 2.5.
The CPT ground state mass also agrees with the value reported in Ref~\cite{kimura_PhysRevC.110.045810_2024}, but is a factor of $\sim$2.5 more precise.
The authors of Ref.~\cite{kimura_PhysRevC.110.045810_2024} did not observe the isomeric state.

\vspace{3mm} \noindent \underline{$A = 155: \; ^{155}$Ce and $^{155}$Pr}
\vspace{2mm}

The AME2020~\cite{ame2020_b_Wang2021} reports an extrapolated mass for $^{155}$Ce, while that for $^{155}$Pr was taken from a CPT \mbox{TOF-ICR} measurement~\cite{van_schelt_prc_2012}.
The low production of $^{155}$Ce, belonging to a $10^{-5}$\% fission branch ~\cite{EnR}, along with its short $t_{1/2}$ of 313(7)~ms~\cite{nubase2020_Kondev2021} limited the precision achieved with the current measurements.
The current CPT results agree with the AME2020 values for both the nuclides, while reporting an eight-times improvement in the precision of $^{155}$Pr.
Both isotopes have been reported in Ref.~\cite{kimura_PhysRevC.110.045810_2024}. The results presented here agree to $\sim$1$\sigma$ with Ref.~\cite{kimura_PhysRevC.110.045810_2024} for both, while reporting uncertainties that are smaller by factors of 5 and 6.5, respectively.


\vspace{3mm} \noindent \underline{$A = 168: \; ^{168}$Gd and $^{168, 168m}$Ho}
\vspace{2mm}

This is the first reported measurement of the $^{168}$Gd mass, with an extrapolated mass being recorded in the AME2020~\cite{ame2020_b_Wang2021}. 
The main limiting factor in the assigned uncertainty here is low statistics as the isotope is produced from a $10^{-5}$\% branch of $^{252}$Cf fission~\cite{EnR}. 
The $^{168}$Ho mass in AME2020~\cite{ame2020_b_Wang2021} is taken from a $\beta$-endpoint measurement~\cite{168Ho_kawade1973decay}, while its isomer mass was calculated from the excitation energy of 59(1)~keV obtained from decay studies~\cite{168Ho_m_chasteler1990decay}. 
The CPT-measured masses for $^{168, 168m}$Ho are $\sim$1.5$\sigma$ lighter than the AME2020~\cite{ame2020_b_Wang2021, nubase2020_Kondev2021} numbers for both the states.
The 58.6(36)~keV excitation energy measured at the CPT agrees with the NUBASE2020 value~\cite{nubase2020_Kondev2021}.
The mass uncertainties reported here are smaller by a factor of 9 than that in the AME2020 for both states.

\vspace{3mm} \noindent \underline{$A = 169: \; ^{169}$Dy and $^{169}$Ho}
\vspace{2mm}

The $^{169}$Dy and $^{169}$Ho masses in the AME2020~\cite{ame2020_b_Wang2021} are adapted mainly from $\beta$-endpoint measurements from Ref.~\cite{169Dy_PhysRevC.42.R1171} and~\cite{169Ho_MIYANO1963315}, respectively. The results reported here agree with the AME2020 values for both the nuclides, with improved precisions of factors $\sim$190 and $\sim$5, respectively. 
Both masses have been measured by the JYFLTRAP double Penning trap \cite{jaries2024probing}, and are around 1$\sigma$ and 5$\sigma$ lighter than the CPT results here. Furthermore, the CPT-measured mass of $^{169}$Dy is about 3.5 times more precise. 
Upon publication of Ref.~\cite{jaries2024probing}, the CPT data for particle identification and measurement of $^{169}$Ho were reexamined, confirming the reported result.
The disagreement indicates a remeasurement of this mass is worth pursuing in the future.

\section{Discussion}\label{subsec:disc}

\subsection{Comparison with mass models}\label{subsec:mass_model}

\begin{figure}
\centering
\includegraphics[width=0.99 \columnwidth]{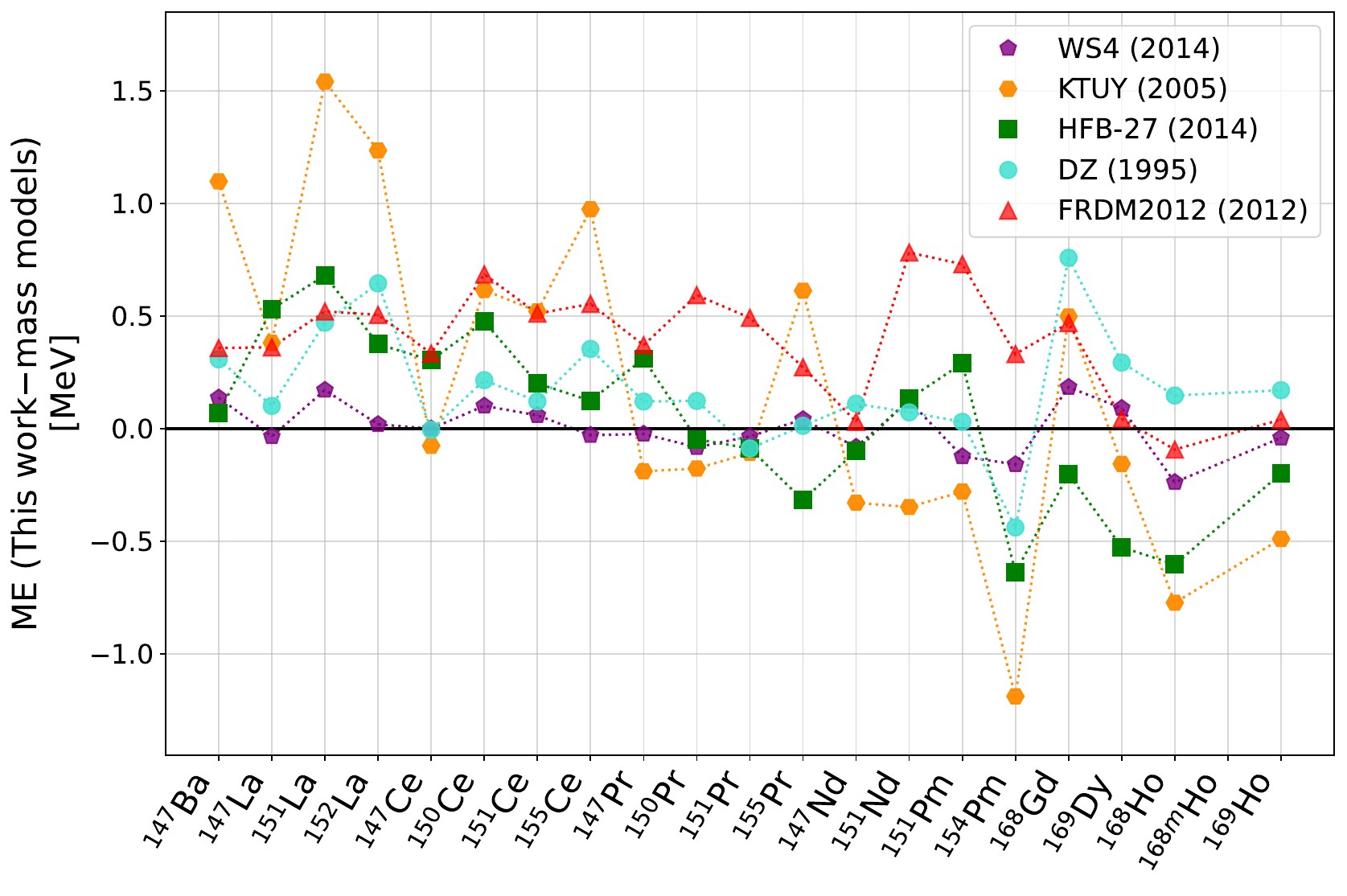}
\caption{Comparison between mass excesses (ME) from this work and five mass models, namely 
WS4~\cite{WS4_2014}, 
KTUY~\cite{KTUY2005}, 
HFB-27~\cite{hfb_27_2014}, 
DZ~\cite{DufloZuker}, and 
FRDM2012~\cite{FRDM2012}.
\label{fig:comparison_results_mass_models}}
\end{figure}

The measured mass results were compared against predictions from five commonly used mass models. 
These include microscopic approaches namely the Hartree-Fock-Bogoliubov (HFB-27)~\cite{hfb_27_2014} and the Duflo-Zuker (DZ)~\cite{DufloZuker} mass models, 
macroscopic-microscopic models like the Weizs{\"a}cker-Skyrme (WS4)~\cite{WS4_2014} and 
the Finite Range Droplet Model (FRDM2012)~\cite{FRDM2012},
and the global Koura-Tachibana-Uno-Yamada (KTUY)~\cite{KTUY2005} mass model \cite{lunney_mass_models_RevModPhys.75.1021}.
The comparison is presented in Fig.~\ref{fig:comparison_results_mass_models} and Table~\ref{tab:mass_comparison} as mass differences between the predictions and measurements of the 20 isotopes persented here.
Additionally, the mass differences between the mass-model predictions and AME2020 results~\cite{ame2020_b_Wang2021, nubase2020_Kondev2021} for all experimentally measured masses are shown in Table~\ref{tab:mass_comparison}.
The differences are presented in the form of absolute mass differences ($\overline{\big| \Delta M \big|}$), and the root mean square mass differences ($\sigma$).
The mass models do not compare well against the current measurements, or the previously measured data from the AME2020~\cite{ame2020_b_Wang2021, nubase2020_Kondev2021}. 
Best agreement is found in the case of the WS4, while the KTUY has the worst agreement with the current measurements. 
A similar trend can be found between the models and the AME2020.
The deviations arise from the theoretical description of the nucleus, which uses a number of free parameters depending on the models. These parameters when fitted to measured masses allow the models to extrapolate better for unknown nuclides~\cite{lunney_mass_models_RevModPhys.75.1021}.
With the development of new advanced rare isotope beam facilities~\cite{frib_article, SAVARD2020258_n126, Fair_Selyuzhenkov_2020, Dilling2014ariel, Kiss_10.1063/5.0036990}, more exotic nuclides will be accessible for precision measurements, enabling better predictions and extrapolations from the models.

\begin{table}[ht]
\begin{ruledtabular}
\caption{Comparison of masses from different mass models with the masses of 20 isotopes reported in this work, and all experimentally measured masses in the AME2020. The differences are expressed in the form of absolute mass differences ($\overline{\big| \Delta M \big|}$), and the root mean square mass differences ($\sigma$). 
\label{tab:mass_comparison}}
\centering
\begin{tabular}{c|c c|c c}

\multirow{3}{*}{Mass models}  & \multicolumn{4}{c}{Comparison, in keV, with} \\ 
  & \multicolumn{2}{c|}{this work} & \multicolumn{2}{c}{AME2020} \\

  & $\overline{ \big| \Delta M \big|}$ & $\sigma$ 
  & $\overline{ \big| \Delta M \big|}$ & $\sigma$ \\ 
\hline

WS4 (2014)~\cite{WS4_2014} & 89 & 109 & 136 & 185 \\
KTUY (2005)~\cite{KTUY2005} & 580 & 712 & 575 & 742 \\
HFB-27 (2014)~\cite{hfb_27_2014} & 311 & 368 & 366 & 508 \\
DZ (1995)~\cite{DufloZuker} & 229 & 308 & 293 & 420 \\
FRDM (2012)~\cite{FRDM2012} & 404 & 460 & 408 & 595 \\

\end{tabular}
\end{ruledtabular}
\end{table}

\subsection{Two-neutron separation energy}\label{subsubsec:S2n}

\begin{figure}[ht]
\centering
\includegraphics[width=0.99 \columnwidth]{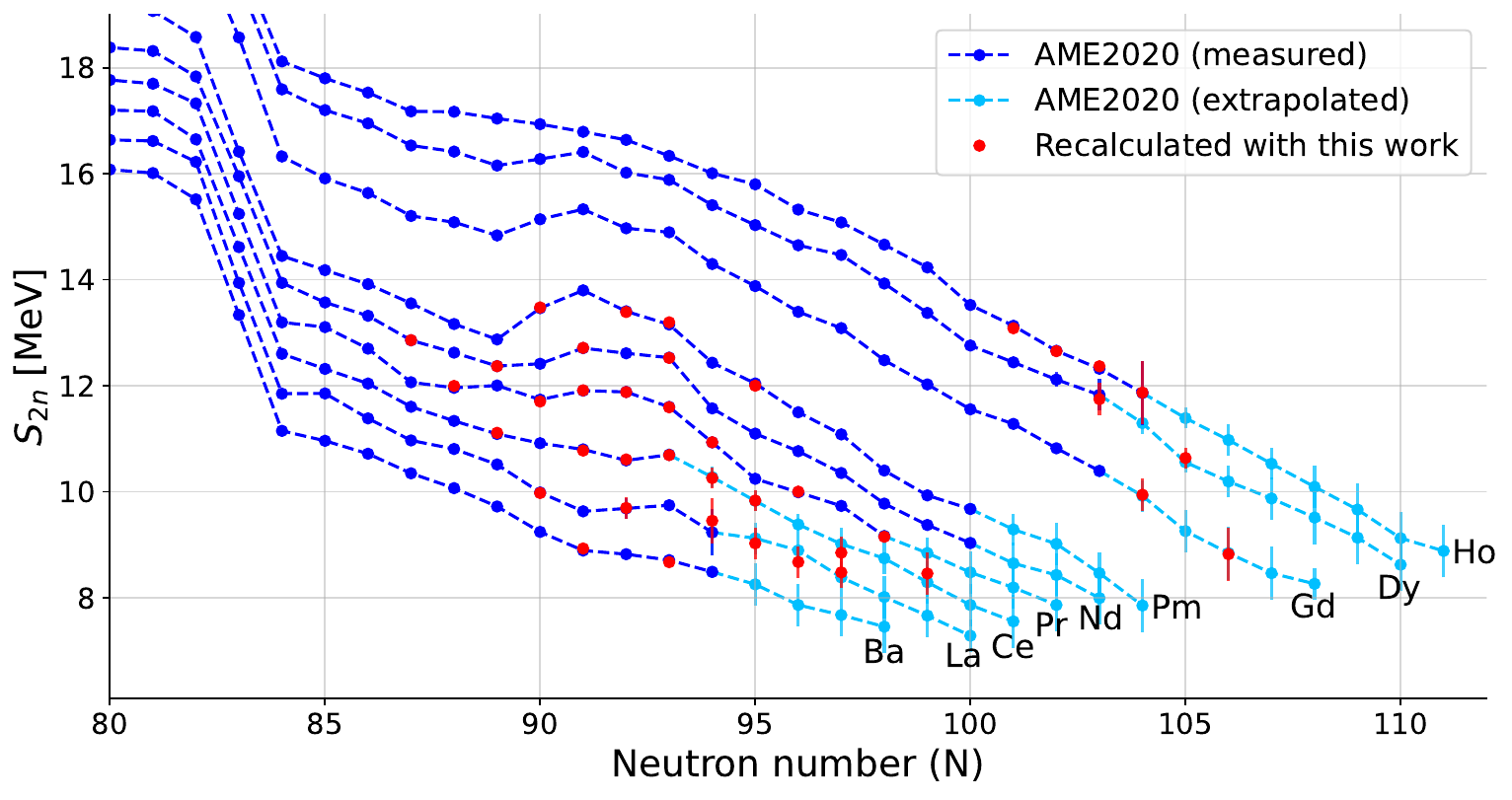}
\caption{Two neutron separation energy curves calculated with mass data from the AME2020~\cite{ame2020_b_Wang2021} and masses presented in this work.
\label{fig:S2n_curve}}
\end{figure}

A key contribution of masses in $r$-process calculations is through the neutron separation energy ($S_{n}$) ~\cite{mumpower_impact_indi_nuc_prop}, given as $S_{n}(Z,N) =  M(Z,N-1) - M(Z,N) + m_{n}$, where $M(Z,N)$ is the mass of a nuclide with $Z$ protons and $N$ neutrons, and $m_{n}$ is the mass of a neutron. 
However, due to the odd-even effects in nuclear binding energies described by the pairing term in the semi-empirical mass formula, the two-neutron separation energy, defined as $S_{2n}(Z,N) =  M(Z,N-2) - M(Z,N) + 2 m_{n}$ can also give useful insights into the local nuclear structure that impacts element formation. The $S_{2n}$ of all the elements whose masses are presented here are calculated using the AME2020~\cite{ame2020_b_Wang2021} values and the masses reported here (see Fig.~\ref{fig:S2n_curve}).

A steep dip in the $S_{2n}$ curve would indicate a shell closure, as can be seen around $N \sim 82$ corresponding to the neutron magic number 82.
On the other hand, a slight positive kink in the curves could imply the presence of some nuclear deformation
\cite{thesis_cpt_orford_phd_long, S2n_positive_kink_deformation_PhysRevLett.87.052502}.
Such kinks can be found over the region $N \sim 89 - 94$ 
in elements Ba$-$Dy with the AME2020~\cite{ame2020_b_Wang2021} masses. 
Calculations made with masses of isotopes of Ba$-$Pm from this work agree fairly well with this trend.

\section{Astrophysical Implications}\label{sec:astro_result}

\begin{figure}[ht]
\includegraphics[width = 8cm]{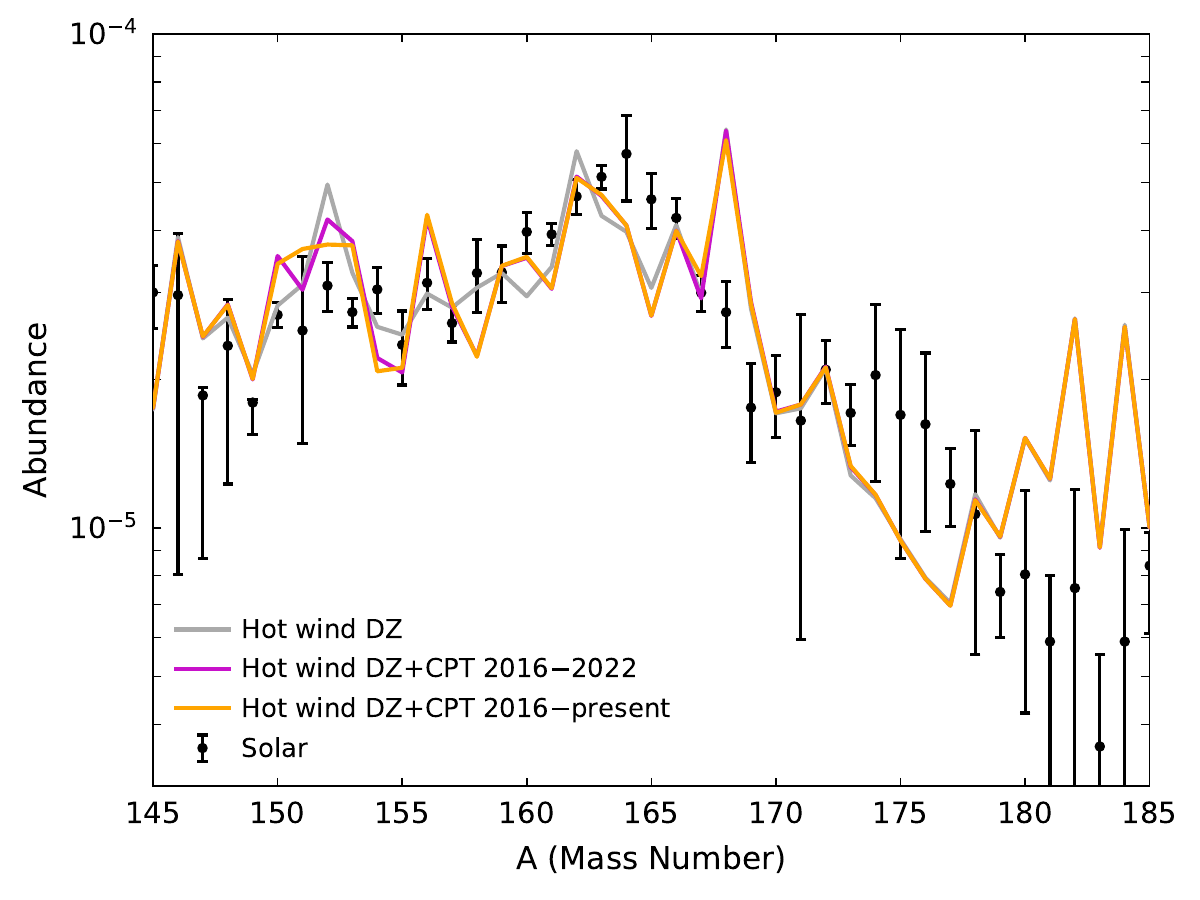}
\caption{Abundance pattern predictions for a hot astrophysical condition of moderate neutron-richness (as in~\cite{VasshMCMC}) given DZ mass prediction across the nuclear chart (grey) then updated with CPT mass measurements. 
The comparison shows results given CPT mass measurements prior to this work (pink) as well as when the new $Z=57$, $Z=58$, and $Z=64$ neutron-rich measurements presented here are also considered (orange). The solar data shown is a symmeterized version of~\cite{Arnould07} as in~\cite{VasshMCMC}.}
\label{fig:abundhotbase}
\end{figure}

The impact of CPT mass measurements on the formation of the $r$-process rare-earth abundance peak was examined by first considering a hot astrophysical outflow with moderate neutron-richness ($Y_e = Y_{p} / (Y_{p}+Y_{n}) = 0.2$) (see Fig.~\ref{fig:abundhotbase}). 
To do so, a baseline was determined from the Duflo-Zuker (DZ) masses~\cite{DufloZuker} and then the CPT masses were applied to inform the neutron capture rates, $\beta$-decay rates, $\beta$-delayed neutron emission probabilities and separation values that enter the nucleosynthesis calculations, as shown in Ref.~\cite{VasshMCMC}. 
Two sets of results were compared: those utilizing just the previous CPT measurements (which includes publications between 2016 and 2022~\cite{orford_prl_Nd-Sm_2018,orford_prc_isomers_2020,thesis_cpt_orford_phd_long, orford_2021_rare_earth2_publshd}) and those with the full CPT measurement set, including the new measurements presented here.
Note AME2016 and AME2020 data was not used in these calculations as the goal was to assess the impact of CPT measurements alone (some of which were reported in AME2020).
Since masses newly probed by this work include $^{152}$La, $^{155}$Ce\footnote{these calculations were conducted before the publication of Ref.~\cite{kimura_PhysRevC.110.045810_2024}} and $^{168}$Gd,
Fig.~\ref{fig:abundhotbase} shows that it is these neutron-rich $Z = 57, 58$ measurements which most impact the abundances, with only very minor differences in the abundance prediction near $A\sim168$. This is because the finalization of abundances mostly takes place below $Z=64$ in the astrophysical conditions considered here. Although the impact primarily comes from incorporating the $Z=58$ mass measurements, note that the $Z=57$ measurement presented here enabled updated predictions for neutron emission probabilities ($P_n$) for $^{151,152}$La. These were previously taken to be $P_0 = 0.85 / P_1 = 0.15$ and $P_0 = 0.74 / P_1 = 0.26$~\cite{Moller03} respectively. 
With the CPT masses, they were found using the BeoH code~\cite{Kawano_Beoh_PhysRevC.78.054601, Mumpower_Beoh_PhysRevC.94.064317, Mumpower_Beoh2_PhysRevC.106.065805} to be $P_0 =0.94 / P_1=0.06$ and $P_0 = 0.56 / P_1 = 0.44$ respectively.
Thus the new values show a decrease in the predicted one neutron emission probability of $^{151}$La
and an increase for $^{152}$La.
Due to the remaining dependence on DZ masses, this calculation does not produce a rare-earth peak consistent with solar data since it is ultimately missing the ability to hold nuclei in place near $A=164$ (\textit{i.e.} cause a `pile-up' in abundances). However it is clear that the suite of CPT masses across the nuclear chart completely change abundance predictions on the left side of the rare-earth peak. Additionally, these measurements begin to encroach upon the region of the nuclear chart which most influences peak formation, evidenced by the differences seen in Fig.~\ref{fig:abundhotbase} near $A=164$.

\begin{figure*}
\centering
\includegraphics[width = 18cm]{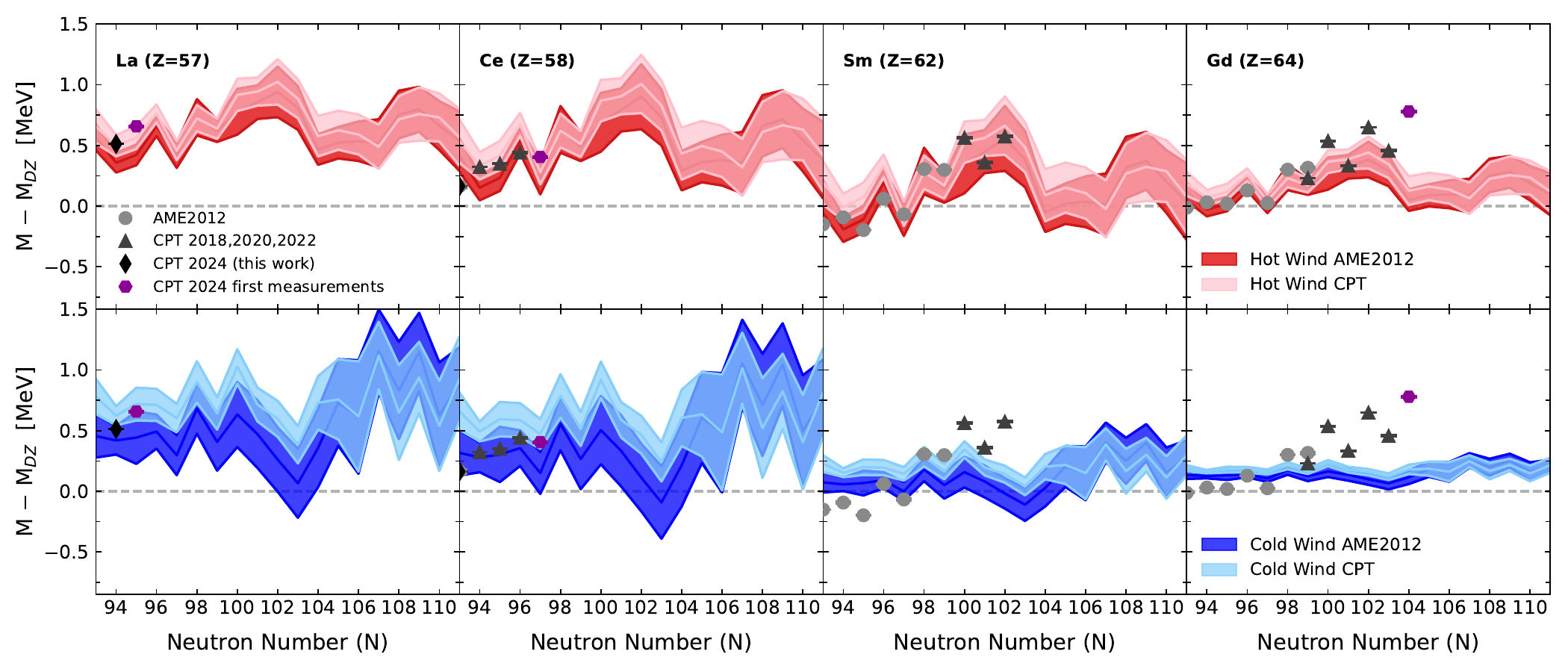}\\
\caption{MCMC mass predictions for the isotopic chains of La (Z=57), Ce (Z=58), Sm (Z=62), and Gd (Z=64) (from left to right) for both a hot astrophysical wind (top row) and a cold wind (bottom row). The darker colored band shows previously published result~\cite{VasshMCMC} which made use of AME2012~\cite{AME2012} masses only to inform the MCMC method. The lighter colored bands show the new results where AME2016 + all CPT measured masses are instead informing the calculations. The previously published results made use of 50 parallel MCMC runs; here the new lighter colored bands are determined from 50 runs in the hot wind case and 35 runs in the cold wind case. Note that the measurements presented in this work are highlighted separately from past CPT published masses, with the cases that are the first precision Penning Trap measurements featured in purple.}
\label{fig:masspred}
\end{figure*} 

\begin{figure}[ht]
\includegraphics[width = 8cm]{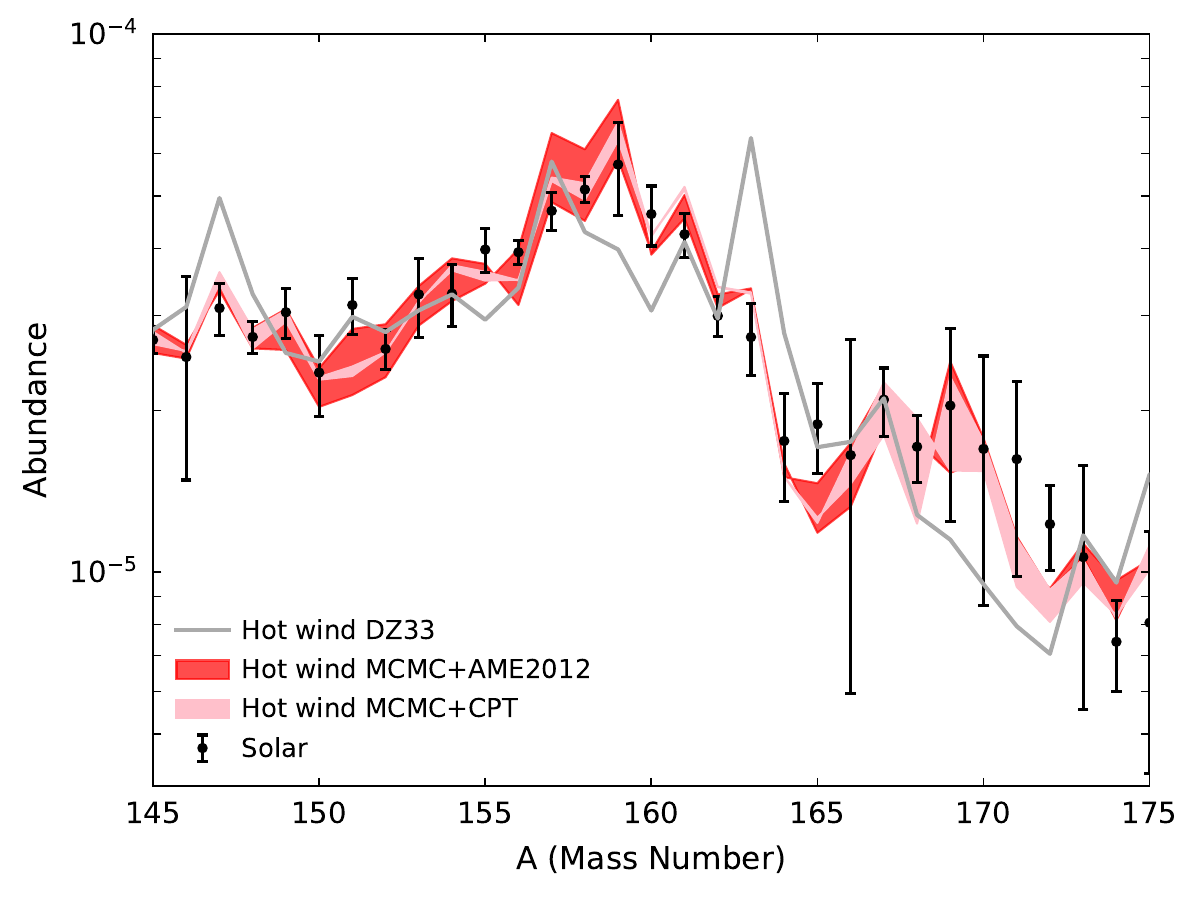}
\hspace{0.25cm}
\includegraphics[width = 8cm]{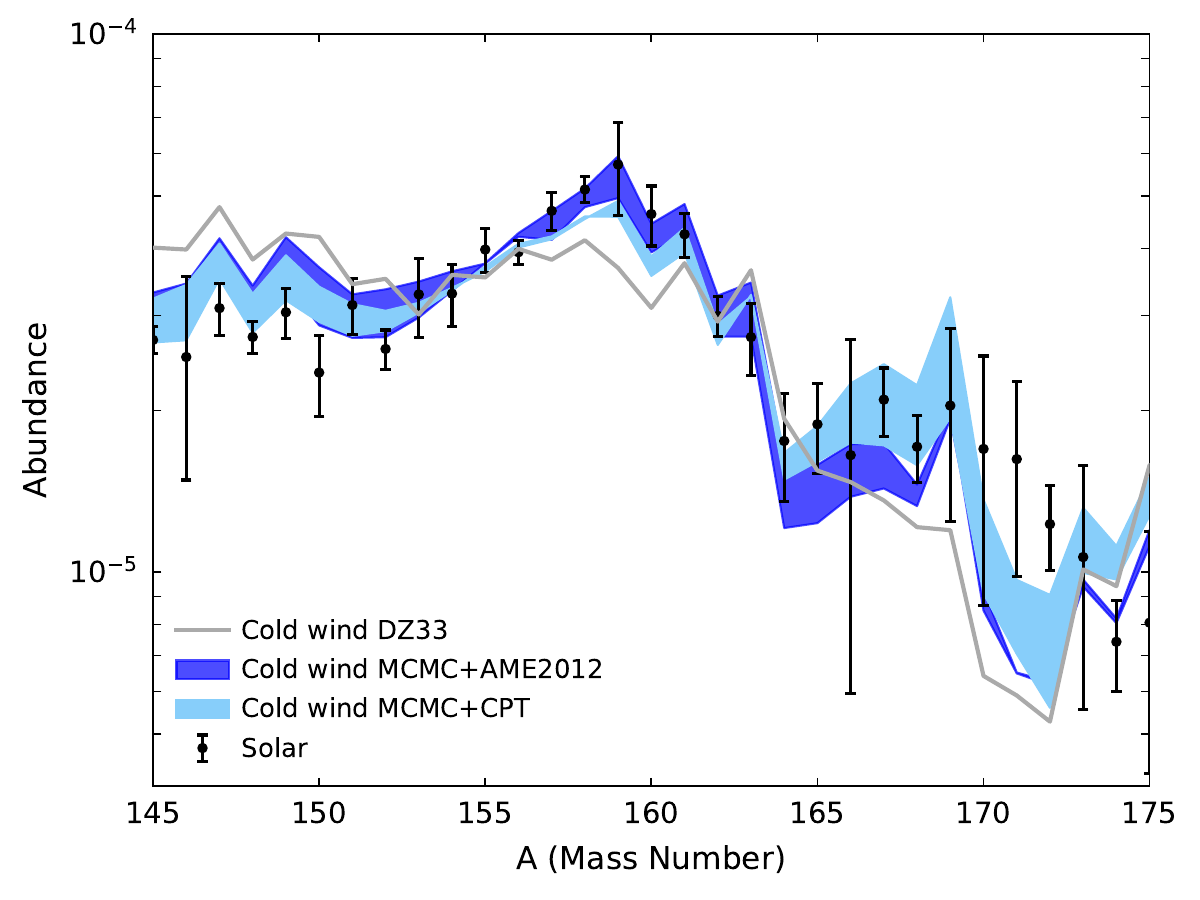}
\caption{Rare-earth peak abundance predictions after applying the masses determined by the MCMC method for the hot wind (top) and cold wind (bottom) conditions considered here. The comparison shows the abundance range found given past MCMC predictions informed by AME2012 (darker band) versus the results given the refined predictions which make use of the CPT mass measurements for neutron-rich rare-earths (lighter band). For reference, the grey line shows the baseline abundance predictions prior to the MCMC adjustments given the DZ mass model.}
\label{fig:abund}
\end{figure}

The abundances in Fig.~\ref{fig:abundhotbase} were then used as a baseline for MCMC calculations which seek to predict the masses capable of forming the rare-earth peak observed in the solar abundances. 
In Fig.~\ref{fig:masspred}, two distinct types of moderately neutron-rich ($Y_e=0.2$) conditions were explored, as can be found in accretion disk winds, \textit{i.e.} the `hot' and `cold' cases in~\cite{VasshMCMC}.  In previous works~\cite{VasshMCMC,VasshSolar,VasshFissMCMC}, the MCMC predictions were not only informed via the $\chi^2$ fit to solar abundances but also constrained by their rms deviation from AME2012 data. 
Those results are replicated here by the darker red and blue bands in Fig.~\ref{fig:masspred}.
The nuclear mass data was updated informing the calculation to use the full suite of CPT masses measured over the last decade which is shown as the lighter red and blue bands in Fig.~\ref{fig:masspred}. 
It can be seen in Fig.~\ref{fig:masspred} that the MCMC method is indeed sensitive to new information through the CPT mass incorporation into $\sigma_{\text{rms}}$. 
In addition, the mass prediction bands were taken from Fig.~\ref{fig:masspred} and propagated 
through to nucleosynthesis calculations as shown in Fig.~\ref{fig:abund}.  
As can be seen in Fig.~\ref{fig:abund}, all scenarios are able to form a rare-earth peak, although the light blue band appears to have the most difficulty hitting the Solar data error bars at the peak.

The most critical mass region for the formation of the rare-earth peak is centered at $Z=60$ for the hot scenario and on $Z=58$ for the cold scenario (see~\cite{VasshMCMC}), and therefore the parameterization used to determine the isotropic mass structure is centered on these elements, $M(Z,N)=M_{DZ}(Z,N)+a_Ne^{-(Z-C)^2/2f}$. The physics of the built-in fall-off, controlled by $f$, is that isotopic chains far from the central element contribute very little toward the formation of the rare-earth peak and this method has little predictive power as to what these masses are.  This manifests as the decreasing thinness of the red and blue bands as one gets further from the central isotopic chains. 
The previous calculations hinted towards hot conditions as those being compatible with both CPT measurements and solar abundance data. This was deduced from the agreement between the hot case mass predictions after $N=100$ and the CPT measurement of $^{164}$Sm ($N=102$). This can be seen in Fig.~\ref{fig:masspred}. 
Keeping in mind that the cold case formation scenario is less sensitive to Sm masses, the cold case is additionally slightly disfavored.

Thus with the hot case of primary interest, the new result was determined from 50 parallel MCMC runs so that the statistics of the new and old results are directly comparable.
For the cold case, 35 MCMC runs were conducted as a test of whether the general mass trend survives. This means the cold case error band here (light blue) could be underestimated by roughly 0.05 MeV, as discussed in~\cite{VasshSolar}, and therefore small discrepancies such as those between the light blue band and the CPT masses at $Z=58$ should not be viewed as significant. Even when informed by extended mass data set provided by the CPT, the cold wind mass trend behavior at $N=100$ remains, 
so measurements of nuclei nearby to and including $Z=58$ and $N=100$ would be informative. 
However, with the available data the hot wind continues to be slightly favored by the MCMC analysis for the reasons discussed earlier.

Additionally, the previous work highlighted the importance of $N=104$ measurements in confirming the hot wind's predicted mechanism of rare-earth peak formation (as highlighted by the dip in the red band of the figure implying this region to be of enhanced stability thereby causing the abundance pile-up that forms the peak). Unfortunately, the $N=104$ measurement presented in this work cannot be used to exclude the hot conditions since peak formation is not significantly influenced by $Z=64$ isotopes, as was demonstrated by Fig.~\ref{fig:abundhotbase}. 
The CPT $Z=64, N=104$ measurement suggests that if there is a nuclear structure feature at $N=104$ responsible for the rare-earth peak, it is localized to proton numbers $Z<64$.

Lastly, the significant increase in precision in the abundance predictions from the refined MCMC predictions is discussed.
In Fig.~\ref{fig:abund}, such refined abundance predictions can be seen to be especially prominent on the left side of the rare-earth peak, reminiscent of the result demonstrated in Fig.~\ref{fig:abundhotbase} that CPT masses are directly influential for $A<164$ abundances. 
However, here it can be seen that the peak at $A=164$ and beyond is also significantly refined given that the bands for the mass predictions have adjusted themselves based on the new CPT mass information at lower neutron numbers. This directly demonstrates the ability of mass measurements across isotopic chains to inform extrapolations and statistical methods seeking to assign mass values based on trends in the nuclear data at lower neutron numbers. The suite of CPT data for $Z=57-64$ measured over the last decade are not only informing nucleosynthesis calculations directly as inputs but also indirectly constraining abundances beyond their direct reach via their influence on extrapolations and predictions for masses at higher neutron numbers.

\section{Conclusions}\label{sec:conclusions}

In this work, new mass measurements in the neutron-rich rare-earth region have been presented, that are important to understand the ultimate origin of the $r$-process rare-earth abundance peak. This work builds off of previous efforts performed with the CPT at CARIBU and analyzes for the first time the impact of the full suite of CPT measurements on statistical methods. The aim was to pin down the astrophysical conditions, consistent with both measured masses and solar abundance data. Although the measurements presented here encroach upon the $N=104$ region that the MCMC method finds to be key for rare-earth peak formation, it has been demonstrated that the isotopic chain of $Z=64$ is unfortunately too near stability to significantly impact rare-earth formation. Since an enhanced stability at $N=104$ from a nuclear property like deformation could be localized to lower mass numbers in the nuclear chart, this indicates that future measurements at lower proton numbers are needed to be carried out at next generation facilities like FRIB~\cite{frib_article}, the N=126 Factory~\cite{SAVARD2020258_n126}, FAIR~\cite{Fair_Selyuzhenkov_2020}, ARIEL~\cite{Dilling2014ariel}, and KISS~\cite{Kiss_10.1063/5.0036990}. 

Nevertheless, this work explicitly demonstrates the impact that pushing the boundaries along neighboring isotopic chains can have on $r$-process predictions, with the CPT mass measurements directly impacting abundance predictions on the left side of the rare-earth peak when incorporated into the nuclear data that determines the nucleosynthesis predictions. Importantly, the additional indirect impact of such measurements, when they are used to inform statistical methods such as MCMC aiming to predict the unknown masses of neutron-rich nuclei, has been demonstrated for the first time. It is through such collaborative efforts between theory and experiment, with dedicated investigations of specific regions of the nuclear chart, that the astrophysical origin of $r$-process elements could be teased out over the next decade.

\begin{acknowledgments} 

This work was performed with the support of US Department of Energy, Office of Nuclear Physics under Contract No. DE-AC02-06CH11357 (ANL), the Natural Sciences and Engineering Research Council of Canada under Grant No. SAPPJ-2018-00028, and the US National Science Foundation under Grant No. PHY-2011890 and PHY-2310059. This research used resources of ANL’s ATLAS facility, which is a DOE Office of Science User Facility. 
N.V. acknowledges the support of the Natural Sciences and Engineering Research Council of Canada (NSERC). Additional support was provided by the U.S. Department of Energy through contract numbers DE-FG02-02ER41216 (G.C.M), DE-FG02-95-ER40934 (R.S.), and LA22-ML-DE-FOA-2440 (G.C.M. and R.S.). R.S. and G.C.M also acknowledge support by the National Science Foundation Grants No. PHY-1630782 and PHY-2020275 (Network for Neutrinos, Nuclear Astrophysics and Symmetries). 
M.~R.~M. acknowledges support from the Directed Asymmetric Network Graphs for Research (DANGR) initiative at Los Alamos National Laboratory (LANL). LANL is operated by Triad National Security, LLC, for the National Nuclear Security Administration of U.S. Department of Energy (Contract No. 89233218CNA000001).
G. C. M. and R.S. acknowledge support through Exascale Nuclear Astrophysics for FRIB (ENAF) a DOE SciDac project. 
This work was also partially supported by the Office of Defense Nuclear Nonproliferation Research \& Development (DNN R\&D), National Nuclear Security Administration, U.S. Department of Energy. This work was performed in part under the auspices of the U.S. Department of Energy by Lawrence Livermore National Laboratory with support from LDRD project 24-ERD02 and under Contract DE-AC52-07NA27344.

\end{acknowledgments}





%

\end{document}